\documentclass[a4paper,12pt]{JHEP3}

\preprint{IPMU11-0154}

\title{Open FRW universes and self-acceleration from nonlinear massive
gravity}

\author{
A. Emir G\"umr\"uk\c{c}\"uo\u{g}lu $\!$\thanks{Email: \texttt{emir.gumrukcuoglu@ipmu.jp}}~, 
Chunshan Lin $\!$\thanks{Email: \texttt{chunshan.lin@ipmu.jp}}~
and Shinji Mukohyama $\!$\thanks{Email: \texttt{shinji.mukohyama@ipmu.jp}}\\
 IPMU, The University of Tokyo, Kashiwa, Chiba 277-8582, Japan}

\date{\today}

\abstract{
 In the context of a recently proposed nonlinear massive gravity with
 Lorentz-invariant mass terms, we investigate open
 Friedmann-Robertson-Walker (FRW) universes driven by arbitrary matter 
 source. While the flat FRW solutions were recently shown to be absent,
 the proof does not extend to the open universes. We find three
 independent branches of solutions to the equations of motion for the
 St\"uckelberg scalars. One of the branches does not allow any
 nontrivial FRW cosmologies, as in the previous no-go result. On the
 other hand, both of the other two branches allow general open FRW
 universes governed by the Friedmann equation with the matter source,
 the standard curvature term and an effective cosmological constant
 $\Lambda_{\pm}=c_{\pm}m_g^2$. Here, $m_g$ is the graviton
 mass, $+$  and $-$ represent the two branches, and $c_{\pm}$ are
 constants determined by the two dimensionless parameters of the theory.
 Since an open FRW universe with a sufficiently small curvature constant
 can approximate a flat FRW universe but there is no exactly flat FRW
 solution, the theory exhibits a discontinuity at the flat FRW limit. 
}

\begin{document}

\section{Introduction}

Gravity remains the most mysterious among the four fundamental forces in
nature. Experimentally, we do not know how gravity behaves at distances 
shorter than $\sim 0.01$ mm or longer than $\sim 1$ Gpc. Thus, it is 
natural to ask whether gravity can be modified at shorter or longer
distances in a theoretically controllable and experimentally viable 
 way. While this question at short distances is relevant for quantum
 gravity, the question at long distances might potentially address the
 mysteries of the universe, such as the cosmological constant problem,
 dark energy and dark matter.

One obvious question associated with modification of gravity at long
distances is ``Can gravitons have a small nonvanishing mass?'' This
question has been investigated in the classical work by Boulware and
Deser~\cite{Boulware:1973my} with a negative conclusion: Einstein
gravity with a nonvanishing mass term exhibits a ghost in nonlinear
level even if the mass term is carefully chosen in the linear level a la
Fierz-Pauli~\cite{Fierz:1939ix}.

Recently a new theory of Lorentz-invariant, nonlinear massive gravity
was introduced~\cite{deRham:2010ik,deRham:2010kj}. Not only the linear
terms but also nonlinear terms at each order are carefully chosen so
that ghost does not show up in the decoupling limit. This theory thus
has a potential to be free from the Boulware-Deser ghost in the fully
nonlinear level~\cite{Hassan:2011hr,deRham:2011rn,deRham:2011qq},
although a different type of ghost within $5$ degrees of freedom of
a massive spin-$2$ field~\cite{Higuchi:1986py} has not been analyzed
yet.

Equipped with a candidate theory of ghost-free massive gravity, it is
natural to study its cosmological implications. Especially, in order to
distinguish this theory from other theories of long-distance
modification of gravity or models of dark energy, an analysis of the 
cosmological perturbations is expected to be useful. Indeed, even if two
different theories give the same cosmic history for the
Friedmann-Robertson-Walker (FRW) background, evolution of perturbations
may differ and act as a discriminator.

On the other hand, in a recent study \cite{D'Amico:2011jj}, it was
argued that the theory contains no nontrivial homogeneous and isotropic
universe (FRW cosmologies). As it was correctly noted and elaborated in
\cite{D'Amico:2011jj}, the absence of FRW cosmologies by itself does not
imply a conceptual or observational problem as long as there are non-FRW
solutions that become more homogeneous and isotropic in the small
graviton mass limit. Nevertheless, this poses a disadvantage, at least
at a technical level: the analysis of the cosmological perturbations
become significantly complicated. For instance, the standard strategy
based on the harmonic expansion should be modified in inhomogeneous or
anisotropic backgrounds.

The main goal of the present paper is to show that the nonlinear,
Lorentz-invariant massive gravity theory allows open FRW
cosmologies, contrary to the no-go result in \cite{D'Amico:2011jj}. 
Although Ref.~\cite{D'Amico:2011jj} states~\footnote{At least in the
current arXiv version (v1).} ``our conclusions on the absence of the
homogeneous and isotropic solutions do not change if we allow for a more
general maximally symmetric $3$-space'', their no-go result actually
does not extend to the open FRW universes. Since an open FRW universe
with a sufficiently small curvature constant can approximate a flat FRW 
universe but there is no exactly flat FRW solution, the theory exhibits
a discontinuity at the flat FRW limit.

The rest of the paper is organized as follows. In Sec.~\ref{sec:setup},
we describe our setup, i.e. open FRW universes in the nonlinear massive
gravity. In Sec.~\ref{sec:constraint-f}, we study the equations of
motion for the St\"uckelberg scalars and find three independent branches
of solutions. In Sec.~\ref{sec:Friedmann-eq} we show that the Friedmann
equation and the dynamical equation are consistent with each other and
that the Friedmann equation includes an effective cosmological constant
of order $m_g^2$, where $m_g$ is the graviton
mass. Sec.~\ref{sec:summary} is devoted to a summary of this paper and
discussions. The paper is supplemented by an Appendix, where we describe
the coordinate transformation from the Minkowski coordinate to the open
FRW chart of the Minkowski spacetime.

\section{Setup}
\label{sec:setup}

In this section, we study the nonlinear massive
gravity~\cite{deRham:2010kj} described by the $4$-dimensional metric
$g_{\mu\nu}$ and scalar fields $\phi^a$ 
($a=0,\cdots,3$), coupled to arbitrary matter source. The role of the
scalar fields $\phi^a$ is to maintain the general
covariance~\cite{ArkaniHamed:2002sp}. By construction, the matter action
$I_m$ is independent of the $\phi^a$ fields. The total action is  
\begin{eqnarray}
 I & = & I_g+I_m, \nonumber\\
 I_g & = & M_{Pl}^2\int d^4x\sqrt{-g}
  \left[\frac{R}{2} 
   + m_g^2( {\cal L}_2+\alpha_3{\cal L}_3+\alpha_4{\cal L}_4) \right]\,,
\end{eqnarray}
where 
\begin{eqnarray}
 {\cal L}_2 & = & \frac{1}{2}
  \left([{\cal K}]^2-[{\cal K}^2]\right)\,, \nonumber\\
 {\cal L}_3 & = & \frac{1}{6}
  \left([{\cal K}]^3-3[{\cal K}][{\cal K}^2]+2[{\cal K}^3]\right), 
  \nonumber\\
 {\cal L}_4 & = & \frac{1}{24}
  \left([{\cal K}]^4-6[{\cal K}]^2[{\cal K}^2]+3[{\cal K}^2]^2
   +8[{\cal K}][{\cal K}^3]-6[{\cal K}^4]\right)\,, 
\end{eqnarray}
and,
\begin{eqnarray}
 {\cal K}^{\mu}_{\nu} = \delta^{\mu}_{\nu}
  - \sqrt{g^{\mu\rho}\eta_{ab}\partial_{\rho}\phi^a
  \partial_{\nu}\phi^b}\,.
\end{eqnarray}
In the above, the squared brackets denote the trace, $\eta_{ab}={\rm
diag}(-1,1,1,1)$ and hereafter, we set $M_{Pl}=1$.

For the physical metric $g_{\mu\nu}$, we consider an open ($K<0$) FRW
universe 
\begin{eqnarray}
 g_{\mu\nu}dx^{\mu}dx^{\nu} & = & 
  -N(t)^2dt^2 + a(t)^2\Omega_{ij}dx^idx^j\,,
  \nonumber\\
 \Omega_{ij}dx^idx^j & = & 
  dx^2+dy^2+dz^2  -\frac{|K|(xdx+ydy+zdz)^2}{1+|K|(x^2+y^2+z^2)}\,, 
\end{eqnarray}
where $x^0=t$, $x^1=x$, $x^2=y$, $x^3=z$; $\mu,\nu=0,\cdots,3$; and
$i,j=1,2,3$. As for the scalar fields $\phi^a$ ($a=0,\cdots,3$), we
adopt the following ansatz, motivated by the coordinate transformation 
(\ref{eqn:min2open}) from the Minkowski coordinates to the open FRW
chart of the Minkowski spacetime:
\begin{eqnarray}
 \phi^0 & = & f(t)\sqrt{1+|K|(x^2+y^2+z^2)}, \nonumber\\
 \phi^1 & = & \sqrt{|K|}f(t)x,\nonumber\\
 \phi^2 & = & \sqrt{|K|}f(t)y,\nonumber\\
 \phi^3 & = & \sqrt{|K|}f(t)z. 
\end{eqnarray}
This leads to the following diagonal form for 
$\eta_{ab}\partial_{\mu}\phi^a\partial_{\nu}\phi^b$. 
\begin{equation}
 \eta_{ab}\partial_{\mu}\phi^a\partial_{\nu}\phi^b
  = -(\dot{f}(t))^2\delta^0_{\mu}\delta^0_{\nu} 
  + |K|f(t)^2\Omega_{ij}\delta^i_{\mu}\delta^j_{\nu}\,, 
  \label{eqn:etadphidphi}
\end{equation}
where a dot represents differentiation with respect to $t$. Since this expression
respects the symmetry of the open FRW spacetime and does not depend on
the physical metric, the ($0i$)-components of the equation of motion for
$g_{\mu\nu}$ are trivially satisfied. Thus, variation of the action with respect to $N(t)$ and $a(t)$ should correctly give all components of the
equation of motion for $g_{\mu\nu}$.

Without loss of generality, we can assume that $\dot{f}\geq 0$, 
$f\geq 0$, $a>0$ and $N>0$, at least in the vicinity of the time of
interest. It is then straightforward to show that 
\begin{equation}
 {\cal K}^0_0 = 1-\frac{\dot{f}}{N}\,, \quad
 {\cal K}^i_j = \left(1-\frac{\sqrt{|K|}f}{a}\right)\delta^i_j\,, \quad
 {\cal K}^i_0 = 0\,, \quad 
 {\cal K}^0_i = 0\,. 
\end{equation}
Thus, up to boundary terms, the gravity action is reduced to the
following form. 
\begin{equation}
 I_g = \int d^4x\sqrt{\Omega}  \left[-3|K|Na-\frac{3\dot{a}^2a}{N}
      +m_g^2 \left( L_2+\alpha_3L_3+\alpha_4L_4\right) \right]\,,
\label{actIg}
\end{equation}
where 
\begin{eqnarray}
 L_2 & = & 3a(a-\sqrt{|K|}f)(2Na-\dot{f}a-N\sqrt{|K|}f)\,, \nonumber\\
 L_3 & = & (a-\sqrt{|K|}f)^2(4Na-3\dot{f}a-N\sqrt{|K|}f)\,, \nonumber\\
 L_4 & = & (a-\sqrt{|K|}f)^3(N-\dot{f})\,.
\end{eqnarray}

\section{Constraint from St\"uckelberg scalars}
\label{sec:constraint-f}

We now investigate the equations of motion for the St\"uckelberg scalars
$\phi^a$. 

Variation of the action (\ref{actIg}) with respect to $f(t)$ leads to
\begin{eqnarray}
 (\dot{a}-\sqrt{|K|}N)
  \left[\left(3-\frac{2\sqrt{|K|}f}{a}\right)
  + \alpha_3\left(3-\frac{\sqrt{|K|}f}{a}\right)
  \left(1-\frac{\sqrt{|K|}f}{a}\right) \right. 
 & &  \nonumber\\
 \left.
   + \alpha_4
   \left(1-\frac{\sqrt{|K|}f}{a}\right)^2
  \right]
 = 0.  &&
\label{eqn:wrtf}
\end{eqnarray}
This equation has three solutions. The first solution,
$\dot{a}=\sqrt{|K|}N$, implies that the physical metric $g_{\mu\nu}$ 
is Minkowski spacetime in the open FRW chart; it is therefore not a
realistic representation of our universe. Reducing the above equation to
remove this solution, we obtain 
\begin{equation}
\left(3-\frac{2\sqrt{|K|}f}{a}\right)
  + \alpha_3\left(3-\frac{\sqrt{|K|}f}{a}\right)
  \left(1-\frac{\sqrt{|K|}f}{a}\right) 
   + \alpha_4
   \left(1-\frac{\sqrt{|K|}f}{a}\right)^2
  = 0\,,
\end{equation}
which is solved by
\begin{equation}
f = \frac{a}{\sqrt{|K|}}X_{\pm}, \quad
 X_{\pm} \equiv \frac{1+2\alpha_3+\alpha_4
 \pm\sqrt{1+\alpha_3+\alpha_3^2-\alpha_4}}
 {\alpha_3+\alpha_4}\,.
\label{eqn:f-sol}
\end{equation}
Note that these two solutions do not exist if $K=0$ is set. This is
consistent with the fact that there is no nontrivial flat FRW
solution~\cite{D'Amico:2011jj}. On the other hand, for $K<0$, these
solutions are well-defined. Furthermore, while $X_+$ is singular in the
limit $\epsilon\to 0$, $X_-$ remains regular in the limit, where
$\epsilon$ is a small parameter counting the order of $\alpha_{3,4}$,
i.e. $\alpha_{3,4}=O(\epsilon)$
\begin{eqnarray} 
 X_+ & = & \frac{2}{\alpha_3+\alpha_4}+ \frac{5\alpha_3+\alpha_4}{2(\alpha_3+\alpha_4)}
  + O(\epsilon), \nonumber\\
 X_- & = & \frac{3}{2}+ O(\epsilon).
\end{eqnarray}

We conclude this section by the consistency of the equations of motion
for the St{\"u}ckelberg fields. Because of the
identity~\cite{Hassan:2011vm} 
\begin{equation}
 \nabla^{\mu}\left(\frac{2}{\sqrt{-g}}
	      \frac{\delta I}{\delta g^{\mu\nu}}\right)
 = \frac{1}{\sqrt{-g}}
 \frac{\delta I_g}{\delta\phi^a}\partial_{\nu}\phi^a, 
\label{stuckid}
\end{equation}
and the triviality of the ($0i$)-components of the metric equation (see
the comment after Eq.~(\ref{eqn:etadphidphi})), the number of
independent equations of motion for the St\"uckelberg scalars $\phi^a$
is one. Thus, the equation obtained by variation with respect to $f(t)$
considered above contains all the nontrivial information.

\section{Friedmann equation and self-acceleration}
\label{sec:Friedmann-eq}

Variation of the action (\ref{actIg}) with respect to $N(t)$ leads to
\begin{equation}
 3H^2-\frac{3|K|}{a^2} 
  = \rho_m + \rho_g, \quad H\equiv \frac{\dot{a}}{Na}
  \label{eqn:N-eq-pre}
\end{equation}
where $\rho_m$ is the energy density of matter fields in the
$I_m$ term of the action, $H$ is the expansion rate defined using the physical time parameter, and 
\begin{eqnarray}
 \rho_g & = & -m_g^2\left(1-\frac{\sqrt{|K|}f}{a}\right)
  \left[3\left(2-\frac{\sqrt{|K|}f}{a}\right) \right.\nonumber\\
 & & \left.
   + \alpha_3\left(1-\frac{\sqrt{|K|}f}{a}\right)
   \left(4-\frac{\sqrt{|K|}f}{a}\right)
   + \alpha_4\left(1-\frac{\sqrt{|K|}f}{a}\right)^2
  \right]\,,
\end{eqnarray}
is the effective energy density contribution arising from the graviton
mass terms. On the other hand, the dynamical equation for the expansion
can be obtained by varying the action with respect to $a(t)$. After
using Eq.~(\ref{eqn:N-eq-pre}), this leads to 
\begin{equation}
 -\frac{2\dot{H}}{N}
  - \frac{2|K|}{a^2} 
  = (\rho_m+p_m) + (\rho_g+p_g),
\label{Hdot}
\end{equation}
where $p_m$ is the pressure contribution from the matter action $I_m$,
and $p_g$ is the effective pressure contribution of the graviton mass
terms. The combination $\rho_g+p_g$ has a relatively simple expression
as 
\begin{eqnarray}
 \rho_g+p_g & = & -m_g^2
  \left(\frac{\dot{f}}{N}-\frac{\sqrt{|K|}f}{a}\right)
  \left[\left(3-\frac{2\sqrt{|K|}f}{a}\right) \right.\nonumber\\
 & & \left.
   + \alpha_3\left(3-\frac{\sqrt{|K|}f}{a}\right)
   \left(1-\frac{\sqrt{|K|}f}{a}\right)
   + \alpha_4\left(1-\frac{\sqrt{|K|}f}{a}\right)^2
  \right]\,.
\end{eqnarray} 

From Eq.~(\ref{stuckid}), we infer that if the constraint (\ref{eqn:wrtf}) is satisfied, the dynamical equation (\ref{Hdot}) brings no new information as a consequence of Bianchi identites and matter conservation. By using the nontrivial solutions (\ref{eqn:f-sol}) of the constraint
(\ref{eqn:wrtf}), Eqs.~(\ref{eqn:N-eq-pre}) and (\ref{Hdot}) reduce to
\begin{equation}
 3H^2-\frac{3|K|}{a^2} 
  = \rho_m + c_{\pm}m_g^2,  \label{eqn:N-eq}
\end{equation}
and
\begin{equation}
 -\frac{2\dot{H}}{N} - \frac{2|K|}{a^2} = \rho_m+p_m,
  \label{eqn:a-eq}
\end{equation}
where
\begin{eqnarray}
 c_{\pm} &\equiv &
  -\frac{1}
  {(\alpha_3+\alpha_4)^2}
  \left[1+\alpha_3\pm\sqrt{1+\alpha_3+\alpha_3^2-\alpha_4}\right]
  \nonumber\\
 & & \qquad\qquad\qquad\qquad\times
  \left[1+\alpha_3^2-2\alpha_4\pm(1+\alpha_3)
   \sqrt{1+\alpha_3+\alpha_3^2-\alpha_4}\right].
  \label{eqn:cpm}
\end{eqnarray}
The first equation (\ref{eqn:N-eq}) is equivalent to the Friedmann
equation for an open universe driven by arbitrary matter (with energy
density $\rho_m$) and the effective cosmological constant 
\begin{equation}
 \Lambda_{\pm} = c_{\pm}m_g^2. 
\end{equation}
For $c_{\pm}>0$, the system exhibits self-acceleration. The second 
equation (\ref{eqn:a-eq}) is consistent with the first equation and as we stated above, does
not lead to any additional conditions, provided that the matter fluid
obeys the ordinary conservation equation
\begin{equation}
 \dot{\rho}_m + 3\frac{\dot{a}}{a}(\rho_m+p_m) = 0.
\end{equation}
Although $c_+$ is singular when $\epsilon\to 0$, $c_-$ remains regular
(and positive) in the limit, where $\epsilon$ is a small parameter
counting the order of $\alpha_{3,4}$, i.e. $\alpha_{3,4}=O(\epsilon)$. 
\begin{eqnarray} 
 c_+ & = & -\frac{4}{(\alpha_3+\alpha_4)^2}-
  \frac{6(\alpha_3-\alpha_4)}{(\alpha_3+\alpha_4)^2}
  -\frac{3(3\alpha_3-\alpha_4)^2}{4(\alpha_3+\alpha_4)^2}
  + O(\epsilon), \nonumber\\
 c_- & = & \frac{3}{4}+ O(\epsilon).
\end{eqnarray}

\section{Summary and discussions}
\label{sec:summary}

In the context of the recently proposed nonlinear massive gravity with
Lorentz-invariant mass terms, we have investigated open FRW universes
driven by arbitrary matter source. We found three independent branches
of solutions to the equation of motion of the St\"uckelberg scalars. One
of the branches forbids any nontrivial FRW cosmologies. On the other
hand, both of the other two branches allow general open FRW universes
governed by the Friedmann equation with the matter source, the standard
curvature term and the effective cosmological constant 
$\Lambda_{\pm}=c_{\pm}m_g^2$, where $m_g$ is the graviton 
mass, $+$  and $-$ represent the two branches, and $c_{\pm}$ are
constants given in (\ref{eqn:cpm}).

As we already pointed out in the Introduction, to distinguish among
long-distance modified gravity theories and dark energy models, an
analysis of cosmological perturbations is of utmost
importance. Different theories may be distinguished by dynamics of
cosmological perturbations even if the FRW background is exactly the
same. The open FRW solutions found in the present paper provides the
working ground for this purpose.

The rank-$2$ tensor 
$\eta_{ab}\partial_{\mu}\phi^a\partial_{\nu}\phi^b$ 
shown in (\ref{eqn:etadphidphi}) respects the symmetry of the open FRW
universe and does not depend on the physical metric $g_{\mu\nu}$. If we
adopt the gauge in which perturbations of $\phi^a$ vanish, the form of
$\eta_{ab}\partial_{\mu}\phi^a\partial_{\nu}\phi^b$ remains 
the same at any order in perturbative expansion. Therefore, in this
gauge, evolution equations for cosmological perturbations fully respect
homogeneity and isotropy at any order. The same conclusion holds in an
arbitrary gauge as far as genuine gauge invariant variables are
concerned.

On the other hand, because of the absence of closed FRW chart in
Minkowski spacetime, there is no way to construct a closed FRW analogue
of (\ref{eqn:etadphidphi}). For this reason, we expect that there is no
nontrivial closed FRW cosmologies.

We note that Ref.~\cite{D'Amico:2011jj} found a special solution in
which the physical metric is of the FRW form, while the stress tensor of
the St{\" u}ckelberg fields is effectively that of a cosmological
constant (while in \cite{Koyama:2011xz, Nieuwenhuizen:2011sq}, exact 
spherically symmetric solutions with self acceleration have been obtained).
However, the tensor 
$\eta_{ab}\partial_{\mu}\phi^a\partial_{\nu}\phi^b$ in this solution
does not respect the symmetry of the FRW universe. As a result, the
contributions of the graviton mass term to the evolution equations of
perturbations are expected to lead to a breaking of the FRW symmetry. In
other words, this solution does not look homogeneous and isotropic if it
is probed by dynamics of perturbations.\footnote{In the framework of the
alternative formulation of \cite{Chamseddine:2011mu}, similar solutions describing flat, open and closed universes have been
found in \cite{Chamseddine:2011bu}.}

While the proof of the absence of nontrivial FRW cosmologies
in \cite{D'Amico:2011jj} applies to flat FRW universes, our solutions
illustrate that it is not valid for open universes. Since an open FRW
universe with a sufficiently small curvature constant can approximate a
flat FRW universe but there is no exactly flat FRW solution, the theory
exhibits a discontinuity at the flat FRW limit. What this implies for
the dynamics of cosmological perturbations in the flat FRW limit of our
solution deserves detailed investigation. This discontinuity may be a
hint of a strong coupling when the contribution from (negative)
curvature becomes negligible, i.e. in the late time evolution with
self-acceleration. On the other hand, a spatial curvature of a percent
level today is consistent with experimental data and would be sufficient
to describe the present universe. Perturbations of the solutions
introduced in the present paper will be discussed in an upcoming work.

\begin{acknowledgments}
 This work was supported by the World Premier International Research
 Center Initiative (WPI Initiative), MEXT, Japan. S.M. also acknowledges
 the support by Grant-in-Aid for Scientific Research 17740134, 19GS0219,
 21111006, 21540278, by Japan-Russia Research Cooperative Program. 
\end{acknowledgments}

\appendix

\section{Open chart of Minkowski spacetime}

The Minkowski metric
\begin{equation}
 ds_0^2 = \eta_{ab}dX^adX^b, \quad \eta_{ab}={\rm diag}(-1,1,1,1)
\end{equation}
can be rewritten in the open FRW form as
\begin{eqnarray}
 ds_0^2 & = & -(\dot{f}(t))^2dt^2 + |K|f(t)^2\Omega_{ij}dx^idx^j,
  \nonumber\\
 \Omega_{ij}dx^idx^j & = & 
  dx^2+dy^2+dz^2  -\frac{|K|(xdx+ydy+zdz)^2}{1+|K|(x^2+y^2+z^2)}, 
\end{eqnarray}
by the coordinate transformation
\begin{eqnarray}
 X^0 & = & f(t)\sqrt{1+|K|(x^2+y^2+z^2)}, \nonumber\\
 X^1 & = & \sqrt{|K|}f(t)x,\nonumber\\
 X^2 & = & \sqrt{|K|}f(t)y,\nonumber\\
 X^3 & = & \sqrt{|K|}f(t)z, \label{eqn:min2open}
\end{eqnarray}
where $K$ ($<0$) is the curvature constant of $\Omega_{ij}dx^idx^j$, and
a dot represents derivative with respect to $t$.

\end{document}